\documentclass[%
 reprint,
 amsmath,amssymb,
 aps,
]{revtex4-2}

\usepackage{graphicx}
\usepackage{dcolumn}
\usepackage{bm}

\begin{document}

\preprint{APS/123-QED}

\title{Dual-recycled interference-based weak value metrology}

\author{Zi-Rui Zhong}
\author{Wei-Jun Tan}
\author{Yue Chen}
\author{Qing-Lin Wu}%
 \email{qlwu@ccnu.edu.cn}
\affiliation{%
\emph{Department of Physics, Central China Normal University, Wuhan 430079, China}
}%

\date{\today}

\begin{abstract}
Weak-value-amplification permits small effects to be measured as observable changes at the sacrifice of power due to post-selection. The power recycling scheme has been proven to eliminate this inefficiency of the rare post-selection, thus surpassing the limit of the shot noise and improving the precision of the measurement. However, the improvement is strictly limited by the system setup, especially the system loss. Here we introduce a dual recycling model based on the interferometric weak-value-based deflection measurement. Two mirrors, the power-recycling mirror and signal-recycling mirror, are placed at the bright and dark port of the interferometer respectively, creating a composite resonator. The results show that both the power and the signal-to-noise ratio (SNR) are greatly enhanced in a wider range of experimental parameters compared to the power-recycling scheme. This work considerably loosens the constraint of the system setup and further explores the real advantage of weak measurement over traditional schemes.

\end{abstract}

\maketitle

\emph{Introduction}. Weak measurement, first introduced by Aharonov, Albert and Vaidman in 1988\cite{1}, has shown significant potential in small effect estimations. In contrast to traditional measurements, the coupling between the system and the probe is slight, permitting small changes in the parameters to be visibly measurable. This amplification comes at the cost of a low detection probability. Therefore, it can be seen that the Fisher information of the parameters to be measured is concentrated in a few collected signals, resulting in the corresponding SNR being almost the same as the SNR in traditional measurements. With these characteristics, weak measurement has been successfully used in the detection of many small physical effects such as the spin Hall effect of light\cite{2,3,4}, optical beam deflection\cite{5}, phase shift\cite{6,7}, Goos-Hänchen shift\cite{8,9}, ultrasensitive velocity\cite{10}, and magnetic resonance\cite{11}. Generally, the weak value can be written as 
$A_w=\left\langle\psi_f\right| \hat{A}\left | \psi_i \right\rangle /\left \langle \psi_f  | \psi_i  \right \rangle $ where $\left | \psi_i  \right \rangle $ and $\left | \psi_f  \right \rangle $
are the pre- and post-selected states of the system, respectively. $A_w$ may be very large if $\left | \psi_i  \right \rangle $ and $\left | \psi_f  \right \rangle $ are nearly orthogonal, resulting in a very large amount of associated parameters. This is refer to as weak value amplification. However, the post-selected probability written as $P=\left | \left \langle \psi_f  | \psi_i  \right \rangle  \right | ^2 $ is a second-order small quantity that causes the signal to being too weak to detect. A solution named joint weak measurement[\cite{12,13} divides the incident light by the Mach-Zehnder interferometer into two paths for detection.
The probability of each path is almost $50\%$, which is a significant enhancement of the detected signal. Another problem is that the SNR is limited by shot noise. The power-recycling technique, initially introduced by Drever\cite{14}, is useful for precision improvement. By placing a partially transmitting mirror at the bright port of an interference to form a resonant cavity, it allows the failed post-selection photons to be repeatedly reselected, thus increasing the Fisher information of the signal to improve the SNR. Another technique is placing a partially transmitting mirror at the dark port of an interferometer. In this technique, which is named signal-recycling \cite{15}, the signal scales linearly with the number of cycles to improve the precision of the detection.

In the power-recycled weak value amplification scheme, all of the input light can be acquired while maintaining the large point shift associated with previous weak value amplification in principle\cite{16,a1,a2}. However, the enhancement is strictly limited by the system loss $\gamma$ and the reflection coefficient of the power recycling mirror\cite{23}. The dual-recycling technique, first introduced by Meers for gravitational wave detection\cite{17}, combines power-recycling and signal-recycling in one Michelson interferometer to enhance the wave signal. It was demonstrated by Strain in an experiment in 1991\cite{18} and successfully applied to Advanced LIGO\cite{19,20}. In this study, we implement the dual-recycling technique into standard weak-value-based metrology to improve the signal and SNR.

\emph{Power recycling}. We first review the interferometric weak-value setup of using the Sagnac interferometer to measure the beam deflection in Ref. \cite{5}. A continuous wave laser with a transverse Gaussian profile $E_{0}(x)=(N^2/2\pi \sigma ^2)exp(-x^2/4\sigma ^2)$, in which $\sigma$ is the modulated length and $N$ is the average number of available photons per unit time, is sent through the interferometer. A Soleil-Bainet Compensator (SBC) controls the phase difference $\phi$ between the clockwise $\circlearrowright$ and counterclockwise $\circlearrowleft$ light. The small transverse momentum is induced by a piezo-driven mirror and can be measured by a split detector placed at the dark port. In this weak-value-based measurement, the system is which-path of light of the interference. We introduce the which-path operator 
$\hat W=\left |\circlearrowright\right\rangle\left\langle\circlearrowright\right| -\left | \circlearrowleft\right\rangle\left\langle\circlearrowleft\right|$ 
defined with two orthogonal circulating states $\left | \circlearrowright  \right \rangle $ and $\left|\circlearrowleft \right\rangle$.
The initial state can be written as $\left | \psi_i  \right \rangle =1/\sqrt2\left [ iexp(i\phi /2)\left | \circlearrowleft\right \rangle +exp(-i\phi/2)\left | \circlearrowright  \right \rangle  \right ]$, and the post-selected state is $\left | \psi_f  \right \rangle=1/\sqrt2\left [ \left | \circlearrowleft  \right \rangle +i\left | \circlearrowright  \right \rangle  \right ]$. Thus, the weak value, which amplifies the kick $k$ for each collected photons, is given by 
$\left \langle \psi_f\right| \hat{W}\left | \psi_i  \right \rangle /\left \langle \psi_f  | \psi_i  \right \rangle=-i\cot \left ( \phi/2 \right )\approx -2i/\phi$. Under the weak value parameter range of $k\sigma \ll \phi/2\ll 1$, the signal-to-noise ratio(SNR) $\mathcal{R}$ is limited by the detected shot noise $N_{det}$: 
\begin{equation}
\mathcal{R}_s=\frac{\left \langle S \right \rangle }{\sqrt{N_{det}}} \approx 2\sqrt{\frac{2}{\pi } }\sqrt{N_{det}}\frac{2k\sigma }{\phi}.  
\end{equation}
As shown in Fig.1, the power recycling technique is used by placing a partially transmitting mirror at the bright port of the interferometer to reuse the photons that return to the laser . In the absence of the cavity, the number of detected photons is $N_{det}=PN$, where $P=\left | \left \langle \psi_f  | \psi_i  \right \rangle  \right | ^2\approx \left ( \phi/2 \right ) ^2 $ is a very small amount so that $N_{det}\ll N$. With the cavity, the detected number can be ideally increased to N, meaning that all incident photons exit the dark port with the amplified deflection\cite{16}. Therefore, the SNR is improved by $2/\phi$ times, which is also a multiple of the quantum limit improvement. Next, we briefly explain why a resonator can allow entire photons to be detected. We consider an initial amplitude $E_0$ that enters a cavity formed by two same partially transmitting mirrors whose reflection coefficient is r and transmission coefficient is p $(r^2+p^2=1)$. With negligible loss, the amplitude of the light in the cavity can be expressed as:
\begin{equation}
E_{cav}=E_0p\left [ 1+r^2e^{i\theta}+(r^2e^{i\theta})^2 +\dots  \right ]=\frac{p}{1-r^2e^{i\theta}}E_0.
\end{equation}
The amplitude of reflected light is
\begin{equation}
E_r=\left ( -r+\frac{p^2re^{i\theta}}{1-r^2e^{i\theta}}  \right )E_0   
\end{equation}
Regarding a resonant cavity, $\theta$ is satisfied with $\theta=2n\pi$, where n is any integer. Thus, $E_r=0$ and $E_{cav}=\frac{1}{p}E_0$. No photons are reflected back to the laser and the gain factor of the signal is $G=\left|E_{cav}\right|^2/\left|E_0\right|^2=1/p^2$.

This calculation of the general model can be used in the power-recycling scheme. The 'system' state $\left |\psi\right \rangle $ is formed by path states $\left | \circlearrowright  \right \rangle $ and $\left|\circlearrowleft \right\rangle$, and the 'meter' state $\left |\varphi\right \rangle $, with the initial pointer for a single photon expressed as $\left \langle x|\varphi_0 \right\rangle=(1/2\pi \sigma ^2)exp(-x^2/4\sigma ^2)$, represents the transverse profile of the beam. Considering the loss caused by optical traversals, we introduce the operator $\hat{L}=\sqrt{1-\gamma}\hat{1}$, where $\gamma$ is the average power loss of one traversal. The spatial filter(SF) placed at the dark port of the interferometer is used to refresh photons in power-recycled weak value metrology, which correct the traverse shift of the beam each pass. \cite{16} calculates the steady state amplitude exiting the detection port, which is given by the sum of amplitutes from all traversal numbers
\begin{equation}
\left|\left.\varphi\right\rangle\right.=\frac{ip\sqrt{1-\gamma}\sin{\left(\phi/2-k\hat{x}\right)}}{1-r\sqrt{\left(1-\gamma\right)P_+}}\left|\left.\varphi_0\right\rangle\right. 
\end{equation}
where $P_+\approx\cos^2{\left(\phi/2\right)}$. Under the impedance matching conditions of $r=\sqrt{\left(1-\gamma\right)P_+}$ and $p\approx\phi/2$, the amplitude of light reflected back toward the laser is 0: 
\begin{equation}
\left|\left.\varphi_r\right\rangle\right.=\left[-r+\frac{p^2\sqrt{1-\gamma}\cos{\left(\phi/2-k\hat{x}\right)}}{1-r\sqrt{\left(1-\gamma\right)P_+}}\right]\left|\left.\varphi_0\right\rangle\right.=0
\end{equation}
By substituting the impedance matching conditions into $\left | \varphi  \right \rangle $, the total number of detected photons is determined to be
\begin{equation}\label{e5}
N_{det}=N\int_{-\infty}^{\infty}{dx\left|\left\langle x\middle|\varphi\right\rangle\right|^2}\approx N\left(1-\frac{4\gamma}{\phi^2}\right).
\end{equation}
Ignoring higher order terms, the detector has an SNR of 
\begin{equation}
\mathcal{R}_p\approx4\sqrt{\frac{2}{\pi}}\sqrt N\frac{k\sigma}{\phi}\left(1-\frac{2\gamma}{\phi^2}\right).
\end{equation}
The factor $2\gamma/\phi^2$ is derived from the filter loss, which is the minimum loss $\gamma_{min}\approx k^2\sigma^2\phi^2/4$ given by the SF. In the absence of this SF, the photons traveling in the cavity have an opposite transversal move, which tends to diminish the detected signal, resulting in a decrease in the weak value amplification. This is called the walk-off effect\cite{6,25}. The filter can eliminate this effect by acting as a projection back onto the initial state with a small adverse impact on the intensity of the signal. When the loss (including the filter loss) $\gamma\ll\phi^2/4$ is small, the SNR is ideally increased by the weak value factor of $2/\phi=1/\sqrt p$ times that of single pass weak measurement, while the weak value amplification does not change.

However, the $1/\sqrt p$ fold improvement of the SNR is almost unreachable in the experiments. The optical loss $\gamma$, which is non-negligible, profoundly affects the gain. In Ref.\cite{23}, under the conditions of $\gamma\approx0.33$ and $\phi/2 \approx0.32$, the intensity of the signal just increases 2.36 times, which can reach 10 if lower loss is provided. In addition, even at negligible loss, optimal region is narrow in the power-recycling mode. For example, when $\phi/2=0.01$, the matched reflection coefficient is approximately $0.99995\pm0.00004$, which is not easy to achieve technically. We will soon show that the dual-recycling scheme can be used to overcome the aforementioned problems to some extent and has more potential in improving the SNR in nearly realistic conditions.

\begin{figure}[t]
\centering
\includegraphics[trim= 0 0 0 0 ,clip, scale=0.4]{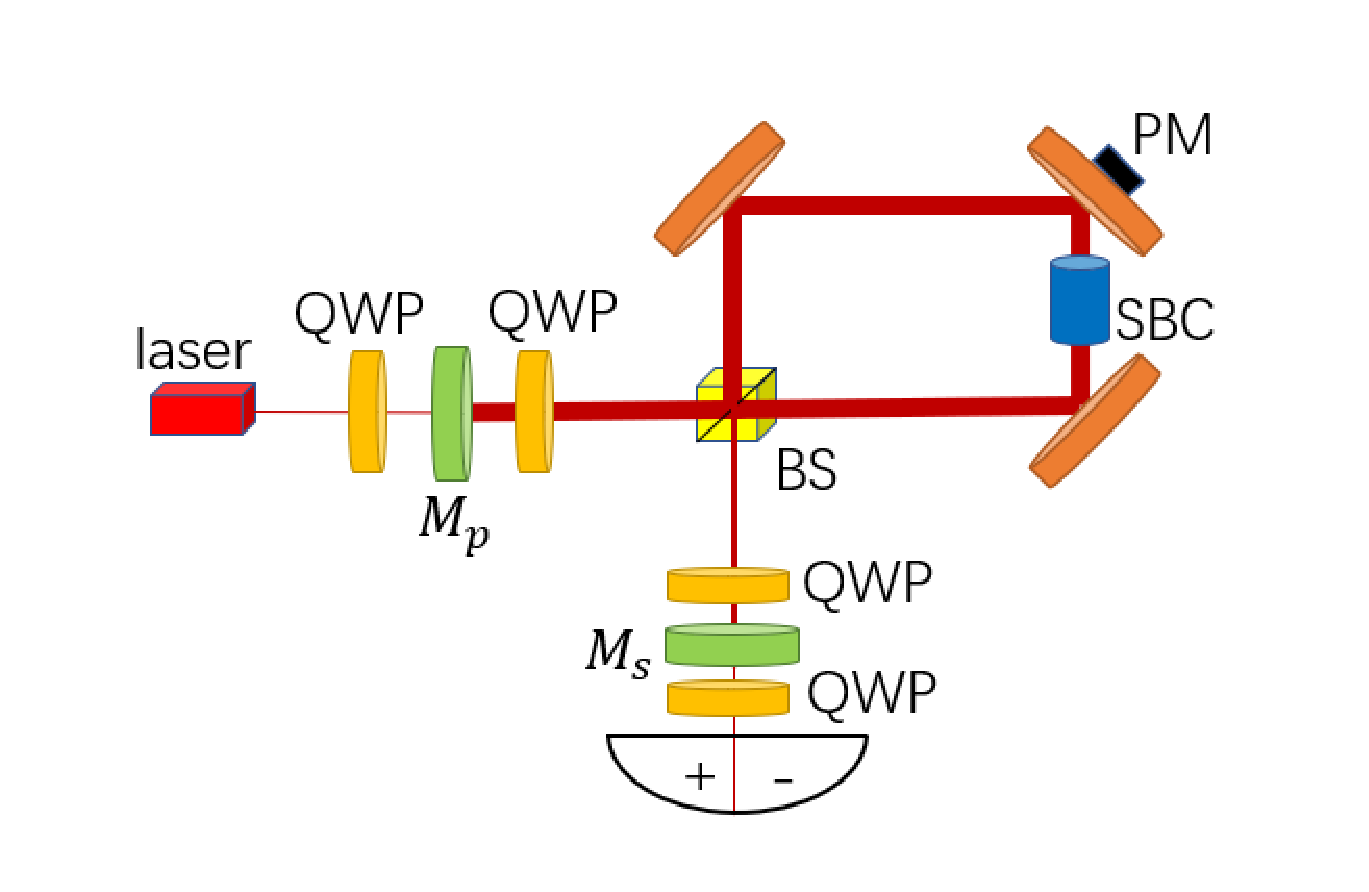}
\includegraphics[trim= 0 0 0 0 ,clip, scale=0.35]{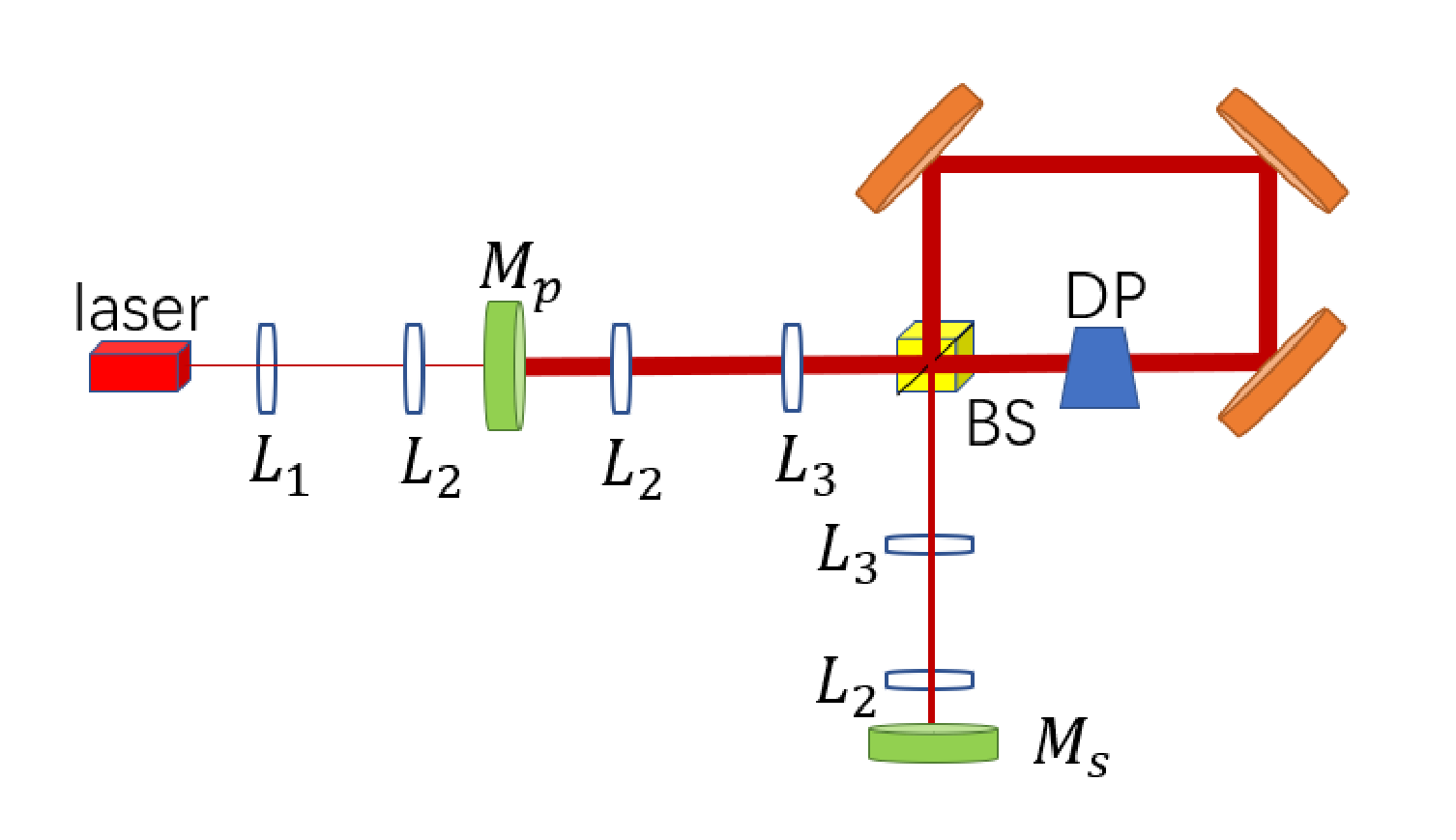}
\caption{(Color online) Schematic of weak-value-based deflection measurement via a dual recycling technique. (a)A continuous Gaussian-shaped wave laser enters to an optical cavity formed by two partially transmitting mirrors$\left(M_p,M_s\right)$ and a Sagnac interferometer. The Soleil-Bainet Compensator (SBC) together with a half wave plate (HWP) controls the phase difference $\phi$ between the two counterpropagating arms. The small transverse momentum is induced by a piezo-driven mirror (PM) and finally measured by a split detector placed at the dark port. (b)Optical mode matching. The lenses $L_1$, $L_2$ and $L_3$ with matched focal lengths are set on proper positions to ensure that the place of the waist and self-reproduction of Gaussian beam are at $M_s$ and $M_p$.
BS: 50/50 beamsplitter, DP: Dove prism}
\label{Fig.1}
\end{figure}

\emph{dual recycling}. Our initial intention for the dual-recycled weak value amplification scheme originated from the feasibility analysis of integrating the power-recycling technique and the signal-recycling technique into one weak measurement setup. Both techniques have their own advantages in improving the $N_{det}$: the partially transmitting mirror placed at the bright port of the interferometer can reflect the photons returned to the laser to the dark port, and all these photons obtain the gain of the resonator at the dark port in turn. It seems that the dual recycling scheme is feasible from a preliminary perspective. However, we need to determine its limitations. When applied to the weak value measurement, the dual-recycling technique cannot break the limit of power recycling because the final number of detected photons $N_{det}$ cannot be larger than the incident photons $N (N_{det}\le N)$ regardless of the design of cavity. Therefore, we pay more attention to its tolerance for $\gamma$ and r, and especially its improvement under nearly realistic loss where we set $\gamma\in[0.1, 0.3]$ in this paper. As shown in Fig. 1, we slightly modify the experimental devices by adding a partially transmitting mirror $M_2$ at the dark port based on the setup in Ref.[16]. To simplify the model, we assume that $M_1,M_2$ are two mirrors with the same parameters and their distances to the BS are equal. In \cite{16}'s power-recycling scheme, The SF is proposed to refresh the transverse profile of cyclic photons, thus eliminating the walk-off effect. This causes a minimum filter loss $\gamma_{min}\approx k^2\sigma^2\phi^2/4$ for ideal optics. However, in this dual-recycling model, the cyclic photons refreshing and transverse shift seem incompatible. We need the SF that refresh all cyclic photons before their last pass through the PM, which is impossible in this two paths, clockwise and counterclockwise, system. Therefore, we remove the SF and consider the walk-off effect in subsequent calculations.

\begin{figure}[t]
\centering	
\includegraphics[trim= 0 0 0 0 ,clip, scale=0.6]{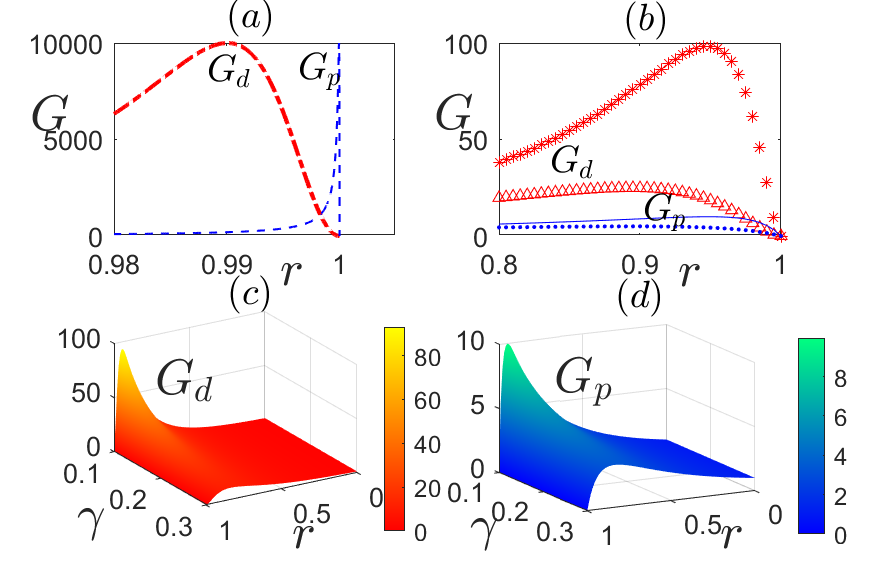}	
\caption{(Color online) Comparison of the power improvement factors of the dual recycling $G_d$ and power recycling $G_p$. (a) $\gamma=0$; $G_d$ (dash-dotted red line) and $G_p$ (dashed blue line) vary with r. (b)$\gamma=0.1,\ 0.2$; $G_d$ and $G_p$ vary with r. The red star, red triangle, solid blue line and blue dot correspond to $G_d (\gamma=0.1)$, $G_d (\gamma=0.2)$, $G_p (\gamma=0.1)$ and $G_p (\gamma=0.2)$ respectively. (c) $G_d$ in 3D. (d) $G_p$ in 3D. $\gamma$: the system loss. r: mirror reflection coefficient. }

\end{figure}

We also consider a transverse Gaussian profile $E_{0}(x)=(N^2/2\pi \sigma ^2)exp(-x^2/4\sigma ^2)$ so that the corresponding initial meter state of a single photon is written as $\left|\left.\varphi_0\right\rangle\right.=\left(1/2\pi\sigma^2\right)^{1/4}exp\left(-x^2/4\sigma^2\right)\left|\left.x\right\rangle\right.$. We introduce two unitary operators ${\hat{U}}_{SBC}=e^{-i \theta\hat{W}/2} $ and ${\hat{U}}_{PM}=e^{ik\hat{x}\hat{W}} $, which represent a net shift $\phi$ produced by the SBC and the weak coupling between the system and meter induced by the PM respectively. The system state of the bright port is defined as $\left|\left.\psi_+\right\rangle\right.=\frac{1}{\sqrt2}\left(\left|\left.\circlearrowright\right\rangle\right.+i\left|\left.\circlearrowleft\right\rangle\right.\right)$ so that the corresponding state of the dark port is described by the orthogonal state $\left|\left.\psi_-\right\rangle\right.=\frac{1}{\sqrt2}\left(\left|\left.\circlearrowright\right\rangle\right.-i\left|\left.\circlearrowleft\right\rangle\right.\right)$. The process of each time the photon entering BS and coming out can be seen as a post-selection projection. Considering that the system state can only be projected onto $\left|\left.\psi_+\right\rangle\right.$ or $\left|\left.\psi_-\right\rangle\right.$, there are four possible projections: 
\begin{equation}
    	\begin{aligned}
M_{11}&=\left\langle\psi_+\middle|{\hat{U}}_{PM}{\hat{U}}_{SBC}\middle|\psi_+\right\rangle=\cos{\left(kx-\phi/2\right)}\\
M_{12}&=\left\langle\psi_-\middle|{\hat{U}}_{PM}{\hat{U}}_{SBC}\middle|\psi_+\right\rangle=i\sin{\left(kx-\phi/2\right)}\\
M_{21}&=\left\langle\psi_+\middle|{\hat{U}}_{PM}{\hat{U}}_{SBC}\middle|\psi_-\right\rangle=i\sin{\left(kx-\phi/2\right)}\\
M_{22}&=\left\langle\psi_-\middle|{\hat{U}}_{PM}{\hat{U}}_{SBC}\middle|\psi_-\right\rangle=\cos{\left(kx-\phi/2\right)}\\
    	\end{aligned}
    \end{equation}
The numbers of Subscripts 1 and 2 represent $\left | \psi_+  \right \rangle $ and $\left | \psi_- \right \rangle$, and Subscript 12 indicates that the initial state $\left | \psi_+  \right \rangle $ is projected into $\left | \psi_- \right \rangle $ in one traversal. 
Therefore, we define a measurement matrix $M$
\begin{equation}
M=\begin{bmatrix}
    M_{11}&M_{12}\\M_{21}&M_{22}\\\end{bmatrix}=\begin{bmatrix}\cos{\left(kx-\phi/2\right)}&i\sin{\left(kx-\phi/2\right)}\\i\sin{\left(kx-\phi/2\right)}&\cos{\left(kx-\phi/2\right)}\\\end{bmatrix}
\end{equation}

$\left(M^n\right)_{12}$ is the change in the state amplitude from $\left | \psi_+  \right \rangle $ to $\left | \psi_-  \right \rangle $ after n traversals, corresponding to the physical process in which incident light starts from the bright port traveling through the cavity n times before reaching the dark port for detection. Therefore, the steady state amplitude detected by the meter is given by the sum of the amplitudes from all traversal numbers,
\begin{equation}
\left|\left.\varphi_d\right\rangle\right.=p\sqrt{1-\gamma}\sum_{n=0}^{\infty}{\left(r\sqrt{1-\gamma}\right)^n\left(M^{n+1}\right)_{12}p\left|\left.\varphi_0\right\rangle\right.}.
\end{equation}
    It is a summation of the convergence series because both $\gamma$ and r are less than 1. Thus, there is a maximum value of n denoted by $n_{max}$, where $(r\sqrt{1-\gamma})^{n_{max}}\approx 0$. The formula above can be simplified as 
\begin{equation}
	\begin{aligned}
    &\left|\left.\varphi_d\right\rangle\right.=p\sqrt{1-\gamma}\sum_{n=0}^{n_{max}}{\left(r\sqrt{1-\gamma}\right)^n\left(M^{n+1}\right)_{12}p\left|\left.\varphi_0\right\rangle\right.}\\
    &=p^2\sqrt{1-\gamma}\left(\frac{M-M(r\sqrt{1-\gamma}M)^{n_{max}} }{I-r\sqrt{1-\gamma}M}\right)_{12}\left|\left.\varphi_0\right\rangle\right.\\
    &\approx p^2\sqrt{1-\gamma}\left(\frac{M}{I-r\sqrt{1-\gamma}M}\right)_{12}\left|\left.\varphi_0\right\rangle\right.\\
    &=\frac{ip^2\sqrt{1-\gamma}\sin{\left(kx-\phi/2\right)}}{1+r^2\left(1-\gamma\right)-2r\sqrt{1-\gamma}\cos{\left(kx-\phi/2\right)}}\left|\left.\varphi_0\right\rangle\right.
	\end{aligned}
\end{equation}
where $I$ is the identity matrix. We perform a Taylor expansion on the function $f\left(x\right)=\sin{\left(kx-\phi/2\right)}/(1+r^2\left(1-\gamma\right)-2r\sqrt{1-\gamma}\cos{\left(kx-\phi/2\right)})$ and make an approximation  $f\left(x\right)\approx f\left(0\right)+xf^\prime\left(0\right)\approx exp\left(-f^\prime\left(0\right)x/f\left(0\right)\right)$ on the condition $2kx/\phi\ll1$ so that the final amplitude of detected state is
\begin{equation}
\left\langle x\middle|\varphi_d\right\rangle\approx A\left(1/2\pi\sigma^2\right)^{1/4}\sin{\left(\phi/2\right)}exp[-\left(x-x_2\right)^2/4\sigma^2],
\end{equation}
where
\begin{equation}
A=\frac{ip^2\sqrt{1-\gamma}}{1+r^2\left(1-\gamma\right)-2r\sqrt{1-\gamma}\cos{\left(\phi/2\right)}}
\end{equation}
and
\begin{equation}
x_2=\frac{2k\sigma^2\cos{\left(\phi/2\right)\left[1+r^2\left(1-\gamma\right)\right]}-2r\sqrt{1-\gamma}}{\sin{\left(\phi/2\right)}\left[1+r^2\left(1-\gamma\right)-2r\sqrt{1-\gamma}\cos{\left(\phi/2\right)}\right]}.
\end{equation}
The corresponding number of detected photons is 
\begin{equation}\label{e1}
	\begin{aligned}
N_{det}&=N\int_{-\infty}^{\infty}{dx\left|\left\langle x\middle|\varphi_d\right\rangle\right|^2}\\
&\approx N\left[\frac{p^2\sqrt{1-\gamma}\sin{\left(\phi/2\right)}}{1+r^2\left(1-\gamma\right)-2r\sqrt{1-\gamma}\cos{\left(\phi/2\right)}}\right]^2
	\end{aligned}
\end{equation}

\begin{figure}[t]
\centering	
\includegraphics[trim= 0 0 0 0 ,clip, scale=0.6]{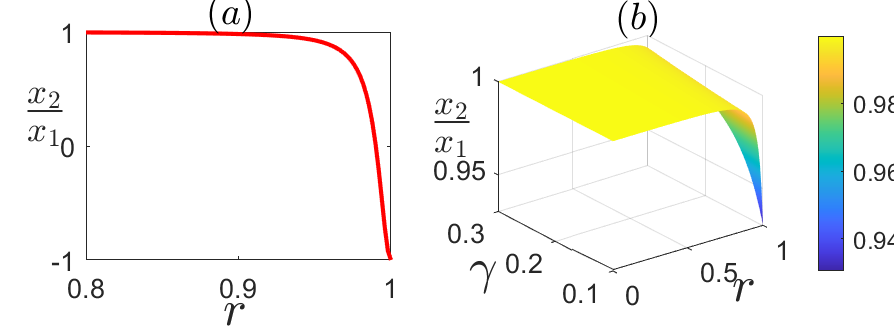}	
\caption{(Color online) The correction of the beam transverse shift based on the walk-off effect. (a) $\gamma=0$; $x_2/x_1$ varies with r. (b) $\gamma\in[0.1,0.3]$; $x_2/x_1$ in 3D. $x_1$ and $x_2$ are the beam transverse shift before and after considering the walk-off effect. $\gamma$: the system loss. r: mirror reflection coefficient.}
\end{figure}

To obtain the maximum $N_{det}$, we need a matched r to make the light reflected back to the laser satisfy $\left | \varphi_r  \right \rangle=0$. Similarly, $\left(M^n\right)_{11}$ represents the change in the amplitude of light reflected back to the laser after n traversals. The meter state amplitude back to the laser is
\begin{equation}
	\begin{aligned}
&\left|\left.\varphi_r\right\rangle\right.=\left[-r+p^2\sum_{n=0}^{n_{max}}{\left(r\sqrt{1-\gamma}\right)^{n+1}\left(M^{n+1}\right)_{11}\ }\right]\left|\left.\varphi_0\right\rangle\right.\\
&=\left\{\frac{p^2\sqrt{1-\gamma}\left[\cos{\left(kx-\phi/2\right)}-r\sqrt{1-\gamma}\right]}{1+r^2\left(1-\gamma\right)-2r\sqrt{1-\gamma}\cos{\left(kx-\phi/2\right)}}-r\right\}\left|\left.\varphi_0\right\rangle\right.
	\end{aligned}
\end{equation}
where the impedance matching condition after neglecting the higher-order terms is
\begin{equation}\label{e3}
r\approx\frac{2-\gamma-2\sqrt{1-\gamma}\sin{\left(\phi/2\right)}}{2\sqrt{1-\gamma}\cos{\left(\phi/2\right)}}.
\end{equation}
By substituting \eqref{e3} into \eqref{e1}, we can obtain the number of detected photons
\begin{equation}\label{e4}
N_{det}\approx N\left(1-\frac{2\gamma}{\phi}\right),
\end{equation}
giving all of them, minus losses. Therefore, the corresponding SNR of the detector is
\begin{equation}\label{e2}
\mathcal{R}_d\approx4\sqrt{\frac{2}{\pi}}\sqrt N\frac{k\sigma}{\phi}\left(1-\frac{\gamma}{\phi}\right).
\end{equation}
It can be seen from \eqref{e4} that $N_{det}$ of dual-recycling reaches a maximum $N$ when $\gamma\ll\phi/2$. $N_{det}$ of power-recycling, given by \eqref{e5}, also reaches a maximum $N$ when $\gamma\ll\phi^2/4$. Regarding the same weak coupling approximation $\gamma\ll\phi/2$, the dual-recycling has wider optimal region of loss.  Considering the experimental loss in practice, the dual-recycling model maintain its advantages. For example, when $\gamma=0.1$ and $\phi/2=0.1$, a common setup, the dual-recycling enables about 40.5 times power improvement with the matched reflection coefficient $r\approx0.90$ while the power-recycling just enables 9.2 times power improvement with a larger matched parameter $r\approx0.95$, where the ideal maximum 
power improvement is $(2/\phi)^2=100$. The similar conclusion is obtained for the SNR analysis.

\emph{$N_{det}$ comparison}. Here we introduce two parameters $G$ and $Q$, which correspond to the power and SNR improvement factor compared to the standard weak value setup, for further comparison.
\begin{equation}
G_p=\left[\frac{p}{1-r\sqrt{\left(1-\gamma\right)}\cos{\left(\phi/2\right)}}\right]^2,
\end{equation}
\begin{equation}
G_d=\left[\frac{p^2}{1+r^2\left(1-\gamma\right)-2r\sqrt{1-\gamma}\cos{\left(\phi/2\right)}}\right]^2.
\end{equation}
where the subscripts $d$ and $p$ correspond to dual-recycling and power-recycling, respectively. In this paper, We set $\phi/2=0.01$, a small angle, to show the large advantages of weak-value-amplification and cavity gain. However, this small post-selected angle corresponds to a high-finesse matched cavity, causing difficulty in controlling experimental accuracy. Thus, in real experiment, proper angle adjustment is necessary. Next, we draw: (1) $G_p$ and $G_d$ vary with the mirrors reflection coefficient r when $\gamma=0$, an ideal loss; (2) $G_p$ and $G_d$ vary with r when $\gamma\in[0.1,0.3]$, common losses.

\begin{figure}[t]
\centering	
\includegraphics[trim= 0 0 0 0 ,clip, scale=0.6]{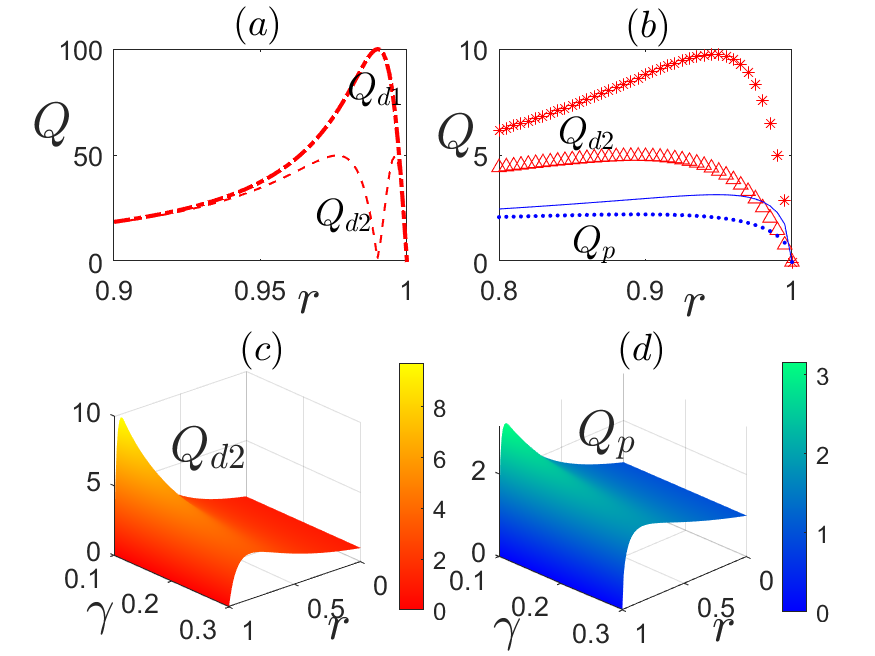}	
\caption{(Color online) The SNR improvement factors of dual recycling ($Q_{d1}$ and $Q_{d2}$) and power recycling ($Q_p$). (a) represents the SNR correction of dual recycling. $Q_{d1}$(dash-dotted line) and $Q_{d2}$(dashed line) vary with r when $\gamma=0$. (b) shows the comparison of the corrected SNR improvement factors between dual recycling and power recycling. 
The red star, red triangle, solid blue line and blue dot correspond to $Q_{d2} (\gamma=0.1)$, $Q_{d2} (\gamma=0.2)$, $Q_p (\gamma=0.1)$ and $Q_p (\gamma=0.2)$ respectively.
(c) $Q_{d2}$ in 3D. (d) $Q_p$ in 3D. $Q_{d1}$ and $Q_{d2}$ are the SNR improvement factors before and after considering the walk-off effect. $\gamma$: the system loss. r: mirror reflection coefficient.}
\end{figure}

Shown in Fig. 2(a), both $G_p$ and $G_d$ can reach the maximum value $\left(2/\phi\right)^2=10000$ as expected. Notably, the mirror reflection coefficients corresponding to the peak values of $G_p$ and $G_d$ are  $r_{pm}$ and $r_{dm}$, respectively. The slope of $G_p$ near the peak value is very large, which means that $r_{pm}$ is limited in a narrow range or the gain effect decreases sharply. $G_d$ changes gently in a wider range of r, making it more tolerant for unavoidable mirror errors. In addition, $r_{dm}$ is approximately 0.99, which is easier to obtain than 0.99995 of $r_{pm}$. From this perspective, the dual recycling scheme with a higher tolerance for r is more advantageous. Figs. 2(b), (c) and (d) intuitively show the comparison of signal amplification effects between the two schemes when $\gamma\in[0.1, 0.3]$. It is obvious that $G_d$ is larger than $G_p$ at the same loss $\gamma$, indicating that the dual recycling scheme is better in overcoming the weak detected signal in the experimental setup.

\emph{The SNR correction}. Due to the walk-off effect, the amplification of the weak value factor involved in the SNR is weakened, resulting in an opposite transverse shift of the Gaussian profile and decreasing the SNR. We can predict that the smaller $\gamma$ is, the larger the walk-off effect is based on the fact that photons travel more times with small losses, which will increase the proportion of the cyclic term $\cos{\left(kx-\phi/2\right)}$. In addition, the walk-off effect in the dual-recycling scheme is larger than that of the power-recycled system. Thus, the SNR described in \eqref{e2}, which is not sufficiently rigorous when considering the non-negligible walk-off effect, should be corrected. In the standard weak value setup, the intensity of the detected light is given by 
\begin{equation}
I_d\approx\frac{N}{\sqrt{2\pi\sigma^2}}\left(1-\gamma\right)\sin^2{\left(\phi/2\right)}exp\left[-\frac{1}{2\sigma^2}\left(x-x_1\right)^2\right],
\end{equation}
where $x_1=-4k\sigma^2/\phi$ is a transverse shift of the Gaussian wave caused by weak interaction, which can be observed by measuring the waveform change.
With the dual recycling cavity, this observable shift changes to $x_2$. The corresponding SNR is then corrected to:
\begin{equation}
\mathcal{R}_c=\frac{x_2}{\bigtriangleup x_2}\approx \sqrt{\frac{2}{\pi \sigma ^2} }  \sqrt {N_{det}} \cdot x_2.
\end{equation}
Similarly, we can obtain the SNR improvement factor in the power recycling scheme:
\begin{equation}
Q_p=\frac{p}{1-r\sqrt{\left(1-\gamma\right)}\cos{\left(\phi/2\right)}}.
\end{equation}
The factors of the dual recycling scheme before and after considering the walk-off effect are noted as $Q_{d1}$ and $Q_{d2}$, which are respectively given by
\begin{equation}
Q_{d1}=\frac{p^2}{1+r^2\left(1-\gamma\right)-2r\sqrt{1-\gamma}\cos{\left(\phi/2\right)}}
\end{equation}
and 
\begin{equation}
Q_{d2}=\frac{p^2}{1+r^2\left(1-\gamma\right)-2r\sqrt{1-\gamma}\cos{\left(\phi/2\right)}} \frac{x_2}{x_1},
\end{equation}
where
\begin{equation}
\frac{x_2}{x_1}=\frac{\phi}{2}\frac{\cos{\left(\phi/2\right)\left[1+r^2\left(1-\gamma\right)\right]}-2r\sqrt{1-\gamma}}{\sin{\left(\phi/2\right)}\left[1+r^2\left(1-\gamma\right)-2r\sqrt{1-\gamma}\cos{\left(\phi/2\right)}\right]}.
\end{equation}

We plot $x_2/x_1$ as a function of r for different losses (1) $\gamma=0$ in Fig. 3(a) and (2) $\gamma\in[0.1, 0.3]$ in Fig. 3(b). As expected, we can obtain a smaller opposite shift with a larger loss. Under ideal lossless conditions, the opposite shift corresponding to r can be of the same magnitude as the transverse shift of weak interactions, and can even reverse the direction of the detected shift. This means that the average number of cycles n of the final probe light to be sufficiently large to allow the cyclic term to dominate is possible. As shown in Fig. 4(a), we compare the SNR before and after correction when $\gamma=0$. This opposite shift results in almost no transverse shift of the detected light under the impedance matching condition ($r_{dm}$). This means that the walk-off effect completely cancels out the weak interaction, leading to the weak-value-amplification effect disappear. In this situation, the noise is infinite and the corresponding SNR is 0. However, this case will not occur because of the existence of losses in the actual system(often between 0.1 and 0.3), where the opposite shift is very small. For example, as shown in Fig.3(b) and Fig. 4(b), (c) and (d) with $\gamma=0.2$ and$\ r=0.9$, the SNR only decreases by less than $0.5\%$, which hardly affects the considerable improvement of the SNR of dual recycling model.

\emph{Loss analysis}. To date, we have just provided the basic calculations of 'unstable' dual-recycling cavity. Generally, the beam will be a diffracting Gaussian beam with a beam waist $w_0$ rather than the parallel beam treated above. Straightly using a flat mirror for beam interference will be exciting multiple modes, leading to the mode loss. Therefore, we need to ensure that the waist size and waist position of the incident laser are the same to that of the resonant cavity itself, forming stable self-reproduction. The partially transmitting mirrors should be curved mirrors or flat mirrors with some matched lenses instead of only flat mirrors. In this paper, we choose the combination of flat mirrors and lenses, which is shown in Fig. 1(b), for a stable cavity configuration. Similar to the previous power-recycling schemes\cite{16}, a Dove prism inside the Sagnac interferometer provides a transverse parity flip to correct the momentum kick and we put the momentum kick instead on the beam splitter for sensitive response. The lenses $L_1$, $L_2$ and $L_3$ with matched focal lengths are set on proper positions to ensure that the place of the waist and self-reproduction of Gaussian beam are at $M_s$ and $M_p$, thus completing the mode matching\cite{23}.

In actual experiments, even if the experimental parameters have been set as required, the length of the resonant cavity is unstable and time-dependent due to the influence of optical platform jitter, temperature, pressure and so on. In \cite{23}, Wang and others proposed using the Pound-Drever-Hall(PDH) system for power-recycling cavity locking: An electro-optic modulator modulates the incident laser into one signal carrier and two sidebands of equal size and opposite phase. By utilizing the influence of cavity length on sideband symmetry, the error signal is output to reflect the change of cavity length and then input into the servo controller for feedback. Next, the feedback currents act on the piezoelectric ceramic to change the length of the cavity, which ensures the stability cavity locking. This PDH system also applies to our dual-recycling scheme, but with some differences. Considering that this is a two-partially-transmitting-mirrors, $M_s$ and $M_p$, dependent cavity, both the paths from $M_s$ and $M_p$ to the beam splitter, noted as $l_s$ and $l_p$, are unstable. A possible way is to adjust the post-selected angle and parameters of $M_s$, $M_p$  to make the light reflected to the laser approximately independent of $M_s$ and use this part of the light to lock the path $l_p$ first. Then add another PDH system to modulate the output light, thus correcting the path $l_s$.

\emph{Conclusion}. In summary, we have proposed a dual recycling scheme to further improve the precision of an interferometric weak-value-based beam deflection measurement compared to the power-recycling scheme. By adding a signal-recycling mirror to the power-recycling cavity, we can similarly acquire all input light to be detected in principle. The results show that the SNR can be greatly improved in a wider range of possible system losses and matched mirror reflection coefficients. However, a small walk-off effect still occurs.

This dual recycling scheme can also be applied to some other experimental realizations because of the ubiquitous problems of probe losses and weak signal exiting in all types of weak value amplification experiments. However, due to the complicated combination of the cavity, the feedback signals require multiple lengths to be adjusted, resulting in the difficult cavity locking, which must be addressed. In addition, We analyze that the filter can be used to eliminate the walk-off effect while leading to the failure of interferometric dual-recycling schemes attributed to the lack of a fixed order of filtering and weak coupling. The polarization-based scheme by using the polarization-beam-splitter may ensure that the light passes through the weak-interaction-site in one direction, thus enabling a filter to refresh photons. In addition, we can combine this technique with some quantum resources such as entangled and squeezed light for further precision improvement in future work.

\emph{acknowledgments}. This work was supported by the National Natural Science Foundation of China (Grants No. 11734015).

\providecommand{\noopsort}[1]{}\providecommand{\singleletter}[1]{#1}%
%

\end{document}